\documentclass[aps,prl,twocolumn,superscriptaddress]{revtex4-1}
\usepackage{graphicx}
\usepackage{color}
\newcommand{\edition}{{}}
\usepackage{amsmath}
\begin{document}

%\preprint{preprint - do not redistribute}

%\title{In-surface electrostatic potential variations in stellarators}
\title{Electrostatic potential variations along flux surfaces in stellarators}


% \address command. The \address command should follow the
% \address can be followed by \email, \homepage, \thanks as well.

\author{M.~A.~Pedrosa, J.~A.~Alonso, C.~Hidalgo, J.~L.~Velasco, I.~Calvo} \affiliation{Laboratorio Nacional de Fusi\'on CIEMAT, 28040, Madrid, Spain}
\author{J.~M.~Garc\'ia-Rega\~na, P.~Helander, R.~Kleiber}
\affiliation{Max-Planck-Institut f\"ur Plasmaphysik, 17491, Greifswald, Germany}
\author{C.~Silva}
\affiliation{Instituto de Plasmas e Fus\~ao Nuclear IST, 1049-001 Lisboa, Portugal}
%\author{M.~A.~Pedrosa} 
%\address{Laboratorio Nacional de Fusi\'on CIEMAT, 28040, Madrid, Spain}
%\author{J.~A.~Alonso}
%\address{Laboratorio Nacional de Fusi\'on CIEMAT, 28040, Madrid, Spain}
%\author{J.~M.~Garc\'ia-Rega\~na}
%\address{Max-Planck-Institut f\"ur Plasmaphysik, 17491, Greifswald, Germany}
%\author{C.~Hidalgo}
%\address{Laboratorio Nacional de Fusi\'on CIEMAT, 28040, Madrid, Spain}
%\author{J.~L.~Velasco}
%\address{Laboratorio Nacional de Fusi\'on CIEMAT, 28040, Madrid, Spain}
%\author{I.~Calvo}
%\address{Laboratorio Nacional de Fusi\'on CIEMAT, 28040, Madrid, Spain}
%\author{C.~Silva}
%\address{Instituto de Plasmas e Fus\~ao Nuclear IST, 1049-001 Lisboa, Portugal}
%\author{P.~Helander}
%\address{Max-Planck-Institut f\"ur Plasmaphysik, 17491, Greifswald, Germany}
%\author{R.~Kleiber}
%\address{Max-Planck-Institut f\"ur Plasmaphysik, 17491, Greifswald, Germany}
%\thanks{}


\date{\today}

\begin{abstract}
First observations of electrostatic potential variations within the flux surfaces of a toroidal magnetic confinement device are presented. Measurements are taken in the TJ-II stellarator with two distant Langmuir probe arrays. The edge floating potentials display differences of several tens of Volts in electron-root wave-heated plasmas. The differences are reduced for higher densities and lower electron temperatures after the ion-root electric field forms at the plasma edge. %\edition{The potential variation derived from first-order neoclassical quasi-neutrality, provides correct order of magnitude differences and tendency with the radial electric field. However, Monte Carlo simulations in the real TJ-II geometry display differences smaller than observed between the angular locations of the probes.}
\edition{Neoclassical Monte Carlo simulations estimate the correct order of magnitude for the overall  variation in potential and predict the trend observed with the radial electric field. However, for the specific location of the probes, the simulations give differences smaller than those observed experimentally.} 
%The understanding of the physical mechanisms responsible for the electrostatic potential variations is particularly important in relation to the radial transport and accumulation of impurities, as these variations cause electrostatic trapping and $E\times B$ radial drifts which can significantly affect radial impurity fluxes.
\end{abstract}

\pacs{52.25.Os,52.30.-q,52.55.Hc}
%\keywords{}

\maketitle

%\section{Introduction}
The accumulation of highly charged impurity ions in the plasma core poses one of the most serious threats to the realization of fusion power production by means of magnetic plasma confinement. In experiments, such impurity accumulation is sometimes observed, but not always, and the theoretical mechanisms of impurity transport are incompletely understood. Due to their high charge, heavy impurity ions are much more sensitive than other plasma particles to variations in the electrostatic potential, and this variation is usually neglected in theories of impurity transport. In this Letter, we present the first experimental observations of the potential variation along the magnetic field in the edge of a toroidal fusion device. In addition, we develop a theoretical prediction thereof, which we compare with the observations, finding values that are large enough to significantly affect the impurity transport. 

The concept of magnetic plasma confinement relies on the construction of a magnetic field structure such that each line of force is contained in a two-dimensional toroidal surface called a \emph{magnetic} or \emph{flux} surface. In the confinement region, magnetic surfaces must exist within the volume enclosed by any one such surface, creating a structure of nested tori. Charged plasma particles can freely explore a flux surface by moving along the field lines, which tends to homogenize the thermal and electrostatic energy over such surfaces. Kinetic plasma theory shows that, if the magnetic field is axisymmetric, the particle distribution function is Maxwellian to zeroth order in a $\delta$ expansion, where $\delta$ is the Larmor radius over the system size \cite{CollisionalHelanderSigmar}. To the same approximation, the electrostatic potential $\phi_0$ is constant on each flux surface, so one can write $\phi_0 =  \phi_0(\psi)$, with flux surfaces defined by $\psi(\mathbf{x}) = \rm{constant}$, and say that $\phi_0$ is a flux function (i.e., it varies in space only through a function $\psi$, called flux). If the magnetic field is not axisymmetric, the distribution function and electrostatic potential are still lowest-order flux functions in most collisionality regimes, but their variation within each surface can be somewhat larger at low collisionality \cite{HoPoF1987}, where the ions are partly electrostatically confined (the so-called $\sqrt{\nu}$-regime). The calculation of the first-order correction to the distribution functions and the electrostatic potential is the subject of the kinetic theory of collisional transport, and it is found that these corrections are not generally flux functions, for the particle cross-field drifts depend on the way the magnetic field strength varies relative to the local direction of the magnetic field vector. 

The parallel variations of the electrostatic energy of singly charged ions and electrons is usually small compared to their thermal energy so that the first-order correction to the electrostatic potential, $\phi_1$, only has a moderate impact on main ion and electron dynamics and transport \cite{MynickPoF1984, BeidlerISHW2005}. However, for impurities of higher charge states $Ze$, ($e$ the elementary charge) the electrostatic energy is $Z$ times larger, whereas their temperatures tend to equilibrate through collisions. Consequently  $\phi_1$ can cause electrostatic impurity trapping in certain regions, in addition to the trapping due to magnetic mirrors. Furthermore, the $E\times B$ radial drift resulting from the $\mathbf{B}\times\nabla\phi_1$ component, can become comparable to the magnetic drifts of impurities because of the $1/Z$ dependence of the latter. Recent simulations have included these effects in the kinetic equation for impurities and shown that the calculated $\phi_1$ importantly affects the predicted radial fluxes of impurities in several magnetic confinement devices \cite{GarciaPPCF2013}. 

Earlier studies in the JET tokamak \cite{IngessonPPCF2000} found indirect evidence of poloidal potential variations caused by radio frequency minority heating, which were postulated to be the cause of the observed poloidal modulation of the nickel impurity density. In recent years, asymmetries in the density and parallel flow of impurities have also been reported from tokamak (see, e.g. \cite{ReinkePPCF2012, ViezzerPPCF2013} and references therein) and stellarator experiments \cite{ArevaloNF2014}. Besides parallel electric fields, inertial forces and friction with main ions in steep gradient regions are considered possible causes of the angular modulation of impurity density. \edition{Several theoretical works have shown the importance of these asymmetries in the radial transport of impurities, both collisional \cite{HelanderPoP1998, LandremanPoP2011} and turbulent \cite{CassonPoP2010, MollenPop2012}.}  

However important, the experimental determination of $\phi_1$ is a complicated task. First, duplicated measurements of plasma potential are not widely available in fusion devices. Second, a precise positioning of the two measuring systems in the magnetic field structure is required to ensure that radial electrostatic potential variations are not misinterpreted as in-surface variations. The measurements reported here were taken with two distant Langmuir probe arrays, whose precise positioning is aided by the simultaneous detection of radially localized zonal-flow-like structures. Significant differences between the floating potential at the two probe locations are observed in electron-root wave-heated plasmas, and are smaller in higher-density, lower-electron-temperature, ion-root plasmas. \edition{Neoclassical Monte Carlo calculations of $\phi_1$ cast overall differences of similar order of magnitudes and trends. However, at the specific locations of the probes, differences in the simulations are generally smaller than those observed experimentally.}

%We show that the order of magnitude and phase of the differences, as well as the observed dependencies on the electric field root, are well reproduced by neoclassical Monte Carlo calculations of $\phi_1$. 
%In this Letter we present the first measurements of electrostatic potential variations within the flux surfaces of a toroidal magnetic confinement device. Measurements are taken with two distant Langmuir probe arrays whose precise positioning is aided by the simultaneous detection of radially localized zonal-flow like structures. Significant differences between the floating potential at the two probe locations are observed in electron-root wave-heated plasmas, that are reduced for higher density, lower electron temperature, ion-root plasmas. We show that the approximate magnitude and phase of the differences, as well as the observed dependencies on the electric field root, are well reproduced by neoclassical Monte Carlo calculations of $\phi_1$. 
%This constitutes the first experimental validation of the neoclassical predictions of the non-constant part of the electrostatic potential.
%\section{Experimental technique}

The experiments were performed in electron-cyclotron-resonance (ECR) or neutral-beam (NB) heated \edition{Hydrogen} plasmas in the TJ-II stellarator ($B$ = 1 T, $\langle R\rangle$ = 1.5 m, $a$ = 0.22 m,  $\iota(a)/2\pi =$ 1.6).  Edge plasma parameters are characterized simultaneously with two similar Langmuir probe arrays located in different toroidal and poloidal sections of the magnetic flux surface. The two arrays are named B and D, according to the toroidal period (named A-D) where they are located. Each of the arrays measures the floating potential in several radial positions simultaneously, spanning about $10\%$ of the minor radius. The magnetic surfaces are calculated with the VMEC magneto-hydrodynamic equilibrium code, and the position of the different probes in the two arrays are calculated from the known positions of the probe actuators. The accuracy of this procedure has been tested by cross-correlating the floating potential fluctuations measured with all the probes. The long-range correlation is due to the existence of zonal-flow-like structures in the plasma edge \cite{PedrosaPRL2008, AlonsoNF2012}. It is found that the maximally correlated pin pairs (one in each of the probes) lie on the same flux surface as calculated with the equilibrium code. Error bars in the flux coordinate of the probes are estimated assuming an uncertainty of $\pm$2 mm in the position of the array along the axis of the actuator.
\begin{figure}
\includegraphics{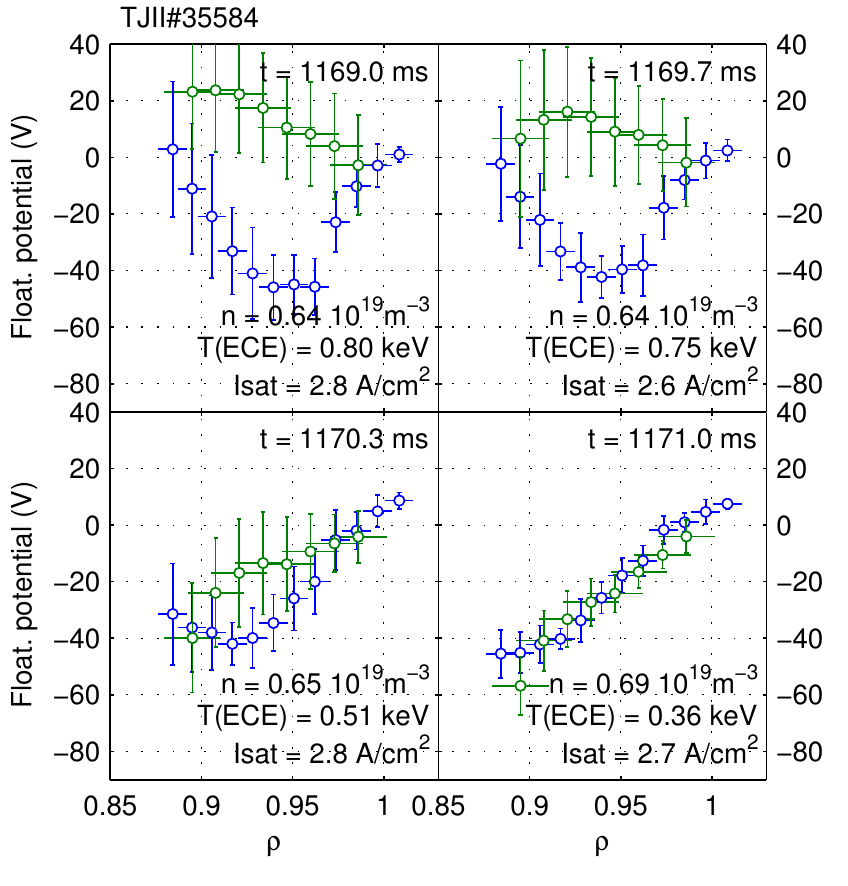}
\caption{\label{fig:ECRoff-prof}Floating potential profiles in the two probe arrays (D-blue, B-green) at four instants during the ECR turn-off (refer to figure~\ref{fig:ECRoff-time}). The profiles are averaged in a 1 ms time window. Vertical error bars correspond to the standard deviation.}
\end{figure}
\begin{figure}
\includegraphics{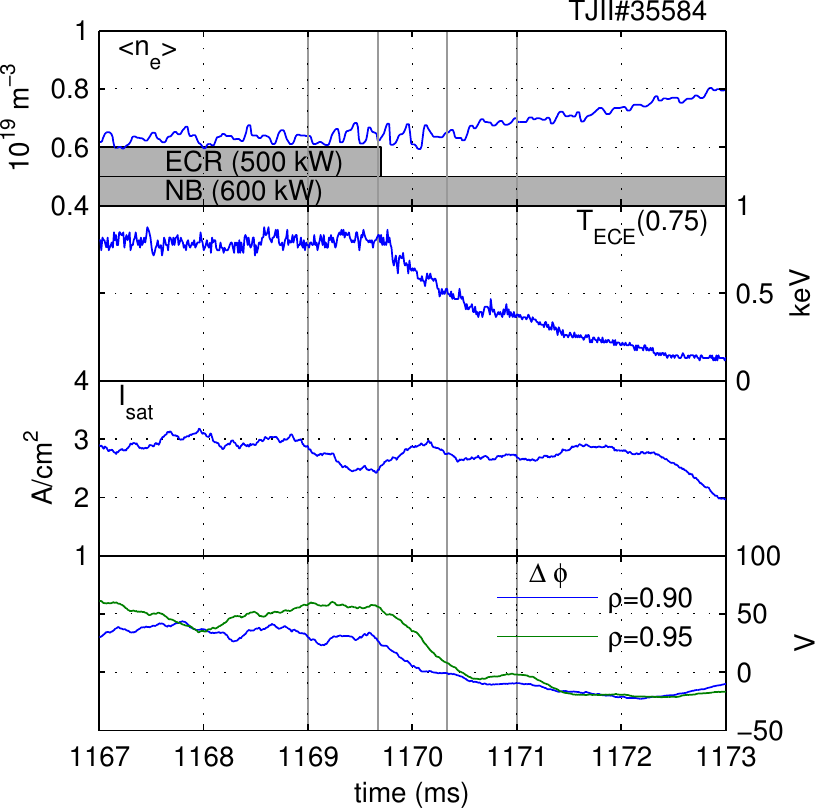}
\caption{\label{fig:ECRoff-time}Evolution of plasma parameters after the turn-off of the ECR heating. Line averaged electron density $\langle n_e\rangle$, electron cyclotron emission temperature $T_{ECE}$ at $\rho = 0.75$, ion saturation current $I_\textrm{sat}\propto n\sqrt{T}$  at $\rho = 0.87$ and floating potential difference $\Delta \phi = \phi_B-\phi_D$ at $\rho = 0.90$ and $0.95$ }
\end{figure}

Differences in the floating potential profiles of several tens of Volts are observed in the edge of the plasma as illustrated in Figure \ref{fig:ECRoff-prof}. The four profiles shown correspond to different instants of a NB+ECR discharge, at about the time of the ECR turn-off (Figure \ref{fig:ECRoff-time}). Differences are large and positive in the ECR phase, go through zero as the electron temperature decreases after the ECR turn-off. The floating potential relates to the plasma potential by an additive term proportional to the electron temperature $T_e$. We assume that the \edition{stationary} $T_e$-variations on flux surfaces are small, so that the in-surface floating potential differences reflect those of the plasma potential.

In order to interpret the observations, we use the first-order quasineutrality condition $n_{i1} = n_{e1} \approx \frac{e\phi_1}{T_{e0}} n_{e0} + O(\delta_e n_{e0})$, where $n_i$ and $n_e$ are the densities of ions (assumed singly charged) and electrons, and $\delta_e$ is the normalized electron Larmor radius. The non-adiabatic part of the electron density ($\sim\delta_e n_{e0}$) is neglected against that of the ions, and therefore
%+
\begin{equation}
\phi_1 \approx \frac{T_{e0}}{e}\frac{n_{i1}}{n_{e0}}~.
\label{eq:qn}
\end{equation}
\edition{Note that this expression implies that $e\phi_1\sim \delta_i T_{e0}$ ($\delta_i$ the normalised ion Larmor radius), whereas electron temperature variations of order $T_{e1}\sim\delta_e T_{e0}$ are expected. This justifies the neglect of stationary $T_e$ spatial variations for the calculation of electrostatic potential differences from the floating potentials.}

The calculation of the ion density perturbation from neoclassical theory requires the solution of the drift-kinetic equation (see e.g. \cite{HoPoF1987})  for the bulk ion distribution function $f_{i1}$, in which $\phi_1$ enters as a source term,
%\begin{equation}\label{eq:dk}
\begin{multline}\label{eq:dk}
(\mathbf{v}_\| + \mathbf{v}_E)\cdot\nabla {f_{i1}} - C^l(f_{i1}) = \\ -\mathbf{v}_M\cdot\nabla\psi \left( \frac{e}{T_{i0}}\frac{d\phi_0}{d\psi}f_{i0}+ \partial_\psi f_{i0}\right) -\mathbf{v}_\|\cdot\nabla\frac{e{\phi}_1}{T_{i0}} f_{i0}~.
\end{multline}
%\end{equation} 
In this equation, the kinetic energy $\varepsilon$ and the magnetic moment are the velocity space variables, $f_{i0} = n_{i0}m_i^{3/2}(2\pi T_{i0})^{-3/2} \exp({-\varepsilon/T_{i0}})$ is the Maxwellian distribution function and $C^l(f_{i1})$ is the linearized collision operator. The parallel, $E\times B$, and magnetic guiding center velocities have subscripts $\|, E$ and $M$, respectively. The neoclassical version of the gyrokinetic Monte Carlo code EUTERPE is used to solve the time-dependent version of this equation with an initial $\phi_1=0$ that is updated from the quasineutrality condition (\ref{eq:qn}) at each time step until a stationary solution is obtained \cite{GarciaPPCF2013}. 

Inspecting equations~\ref{eq:qn} and \ref{eq:dk} reveals some general features of the solution $\phi_1$: a) neoclassically optimized configurations with reduced radial magnetic drifts will exhibit relatively small electrostatic potential variations; b) higher electron temperature generally requires larger potential variations for the adiabatic electron density to balance the ion density variation; c) a positive radial electric field $-d\phi_0/d\psi > 0$ adds to the logarithmic gradients of the mean density and temperature (normally directed inwards) resulting from the $\partial_\psi f_{i0}$ term in Eq.~(\ref{eq:dk}), to produce a larger source term, and thus larger $n_{i1}$ and $\phi_1$. Furthermore, in a stellarator device, the $E\times B$ advection term $\mathbf{v}_E\cdot\nabla f_{i1}$, causes particle trapping/detrapping and may alter the angular dependence of $\phi_1$. 

\begin{figure}
\includegraphics{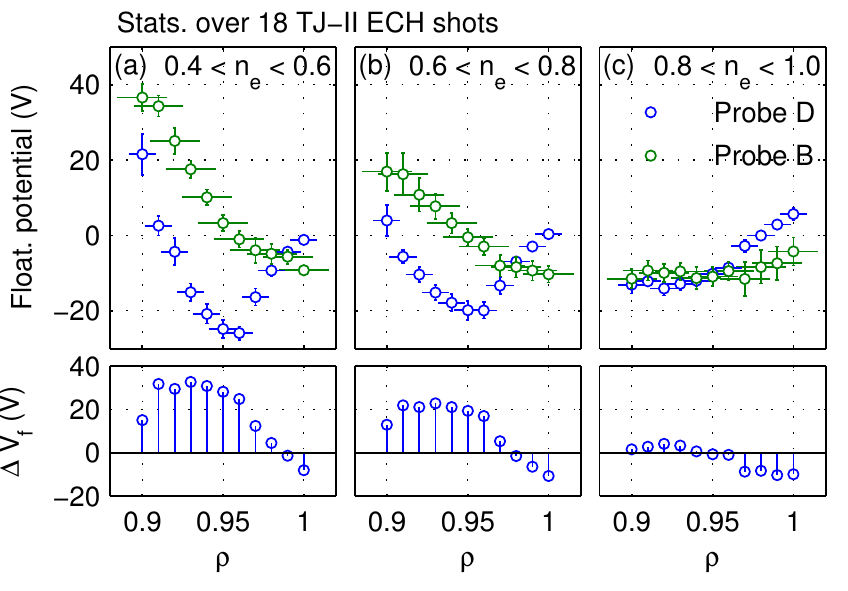}
\caption{Floating potential profiles at the two probe locations averaged in three different line-averaged electron density ranges. Respectively, (a), (b) and (c) correspond to typical electron root, root transition, and ion root conditions. The profile difference is shown in the lower axis. Density ranges are expressed in $10^{19}$m$^{-3}$ units.\label{fig:nscan}}
\end{figure}

\begin{figure}
\includegraphics{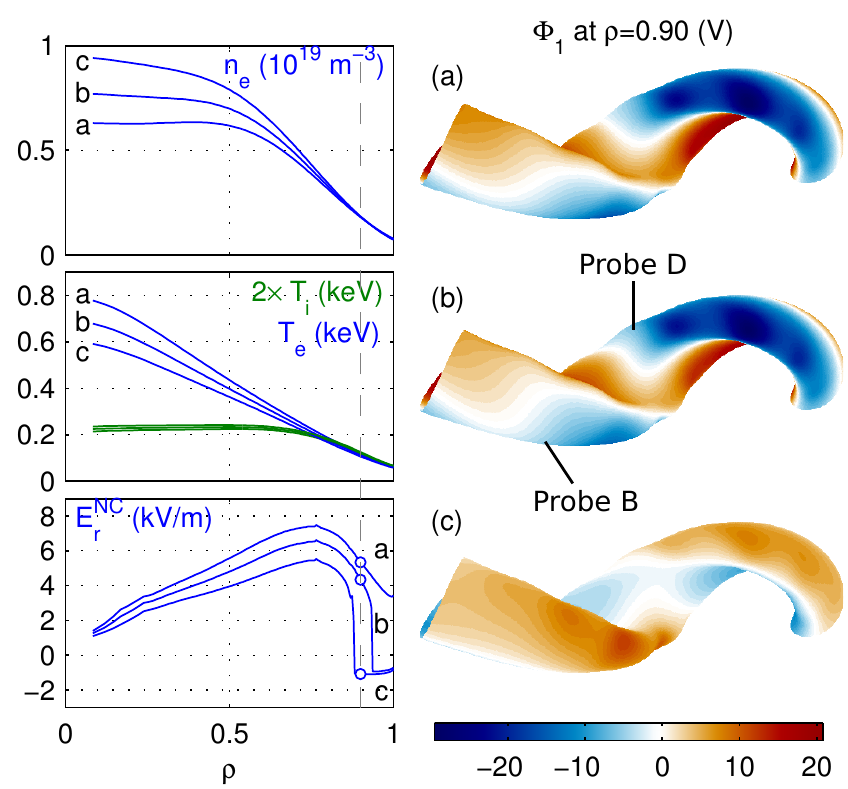}
\caption{Left: Typical profiles of density $n_e$, electron and ion temperatures, $T_e, T_i$, representative of the situation (a) to (c) in figure \ref{fig:nscan}. 
The radial electric field is obtained form the ambipolar condition for the neoclassical fluxes calculated with DKES (see e.g. \cite{VelascoPRL2012}). The three sets of profiles are used as input for the EUTERPE simulations (right and figure \ref{fig:comp}). Right: simulated $\Phi_1$ for the surface $\rho = 0.9$ and plasma parameters shown on the left. \label{fig:sim}}
\end{figure}

\begin{figure}
\includegraphics{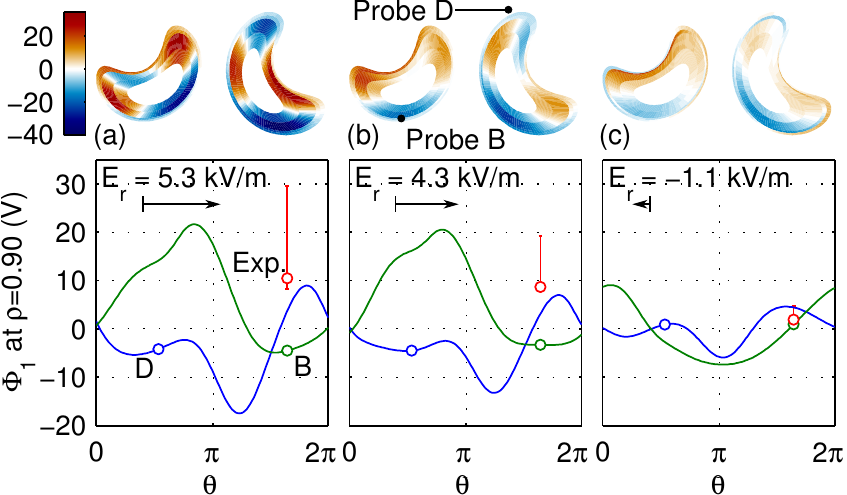}
\caption{Simulation of $\phi_1$ for the three profile sets in figure \ref{fig:sim}. Colormaps for the toroidal sectors of the two probe arrays are drawn at the top, and the poloidal location of the probes is indicated. The poloidal angular variations of $\phi_1$ at radius $\rho=0.9$ for the two sectors is shown at the bottom plots. Green and blue circles labeled B and D mark the simulated value at the location of the two probes. The experimentaly observed differences (see figure \ref{fig:nscan}) are ploted \label{fig:comp} in red over the simulated value B, taking the value at D as reference. Black arrows indicate the magnitude and $\theta$-direction of the $E_r\times B$ flow.}
\end{figure}

\edition{
The tendency observed in figures \ref{fig:ECRoff-prof} and \ref{fig:ECRoff-time} is in agreement with the statistical analysis of the 18-shots database of ECRH plasmas shown in figure \ref{fig:nscan}. This figure shows the floating potential profiles of the two arrays and their difference, conditionally averaged for different line-averaged electron densities around the root transition. The transition from a positive, electron-root radial electric field to a negative ion-root one has been extensively characterized experimentally (see e.g. \cite{PedrosaPPCF2005}), and shown to occur with a relatively minor increase (decrease) in the local electron density (temperature) \cite{VelascoPRL2012}. The root jump is due to the sensitivity of the radial electron flux to the collisionality for those conditions.

%The TJ-II stellarator offers optimum conditions to test the radial electric field dependence. The transition from a positive, electron-root radial electric field to a negative ion-root one has been extensively characterized experimentally (see e.g. \cite{PedrosaPPCF2005}), and shown to occur with a relatively minor change in plasma profiles \cite{VelascoPRL2012}. The root jump is due to the sensitivity of the radial electron flux to the collisionality for those conditions. 
Archetypal TJ-II profiles displaying this electric field root transition at the edge are shown in  figure \ref{fig:sim} (left). These are used as input for the EUTERPE $\phi_1$ calculations shown in figure \ref{fig:sim} (right) for the edge flux surface $\rho= 0.90$, and for several flux surfaces in figure \ref{fig:comp} (top). Consistent with the theoretical discussion following equations \ref{eq:qn} and \ref{eq:dk} and the experimental observations (figure \ref{fig:nscan}), simulations show that the simultaneous decrease of the positive radial electric field and electron temperature when approaching the transition, causes a reduction of the peak values of $\phi_1$. Additionally, the phase of the perturbation changes very noticeably after the root transition in the direction of the difference in the $E\times B$ flow (figure \ref{fig:comp}).}
\edition{%
The detailed comparison of the measured and simulated differences between the angular locations of the two probes is shown in Fig.~\ref{fig:comp}. The surface $\rho = 0.90$ has been chosen for the comparison, although similar results are obtained in the range [0.90, 0.95]. Simulations display small differences between the probes (marked B and D) for three cases. This is in contrast with the experimental values plotted over the B probe, D taken as reference. The sensitivity of the simulations to the experimental errors in the radial and angular location of the probes and/or to the value of $E_r$	 cannot resolve the observed discrepancy. It should be noted that the large values of $e\phi_1/T_{e0}\lesssim 0.5$ at the edge result from large deviations in the ion distribution function, which approaches the validity limits of the perturbative drift-kinetic treatment. The extension of this comparison to the core region will be the subject of future work.} 
%
%\begin{figure}
%\includegraphics{Stats-Paper.pdf}-
%\caption{Statistics of floating potential differences at $\rho=0.93$ for a set of ECR discharges. (a) probability distribution function of the differences. (b)-(c) Expected values of line averaged density and floating potentials conditioned to a certain value of the potential difference\label{fig:stats}.}
%\end{figure}
%

In conclusion, we have reported the first direct experimental observations of in-surface electrostatic potential differences in a toroidal magnetic confinement device. Differences of several tens of Volts are consistently measured in the floating potential values recorded at two different angular locations on the same flux surface. 
\edition{%
These differences are large enough for the electrostatic trapping of impurities to dominate over the magnetic mirror. The ratio of parallel electric field to mirror acceleration, $a_{\phi_1}/a_m\sim (Ze\Delta\phi_1/T_z)(B/\Delta B)$, is estimated to be about $10$ for these conditions ($Z=4, T_z = 60$ eV, $\Delta B / B = 0.25$).
The differences are observed to be reduced for lower electron temperatures and/or more negative radial electric fields. This is consistent with the expected dependencies of the electrostatic potential variation $\phi_1$, derived from first-order neoclassical quasi-neutrality. Monte Carlo simulations of $\phi_1$ in the real TJ-II geometry display peak-to-peak values of the order of magnitude of the experimental observations. However, for the specific location of the probes, the simulated differences are smaller than those observed experimentally.

To improve our understanding of the parallel variation of electrostatic potential is to be deemed an important task in magnetic confinement fusion research, for its effect on and/or combination with impurity density variations can give rise to substantial changes of radial impurity fluxes}. This variation is candidate to explain some of the impurity confinement anomalies observed in stellarators \cite{HirschPPCF2008, YoshimuraNF2009}.

%The authors wish to thank R. Kleiber (IPP) for helpful discussions. 
Simulations presented here were partly performed in the HELIOS supercomputer system (IFERC-CSC, Aomori, Japan).This work was supported by EURATOM and carried out within the framework of the EUROfusion Consortium. This project has received funding from the European Union's Horizon 2020 research and innovation programme. The views and opinions expressed herein do not necessarily reflect those of the European Commission.
%\subsection{Description of the TJ-II probe system and radial callibration procedure}
%\begin{figure}
%\includegraphics{Callibration.pdf}
%\caption{Positioning of the probes in the radial flux coordinate: comparison of magnetic surface mapping and long range correlations between the two probes. The maximally correlated pair of pins always appear to be on the same flux surface as calculated with the vacuum magnetic field. Discharges from three different magnetic configurations are shown. Error bars correspond to a $\pm 2$mm uncertainty in the position of the pins along the direction of the probe drive}
%\end{figure}

%\bibliographystyle{unsrt}
%\bibliography{/home/arturo/Documents/MyOwn/MyNewBib}

\end{document}